# More on SU(3) lattice gauge theory in the fundamental–adjoint plane[*]

Urs M. Heller[a]

[a]SCRI, The Florida State University, Tallahassee, FL 32306-4052, USA

We present further evidence for the bulk nature of the phase transition line in the fundamental–adjoint action plane of SU(3) lattice gauge theory. Computing the string tension and some glueball masses along the thermal phase transition line of finite temperature systems with $N_t = 4$, which was found to join onto the bulk transition line at its endpoint, we find that the ratio $\sqrt{\sigma}/T_c$ remains approximately constant. However, the mass of the $0^{++}$ glueball decreases as the endpoint of the bulk transition line is approached, while the other glueball masses appear unchanged. This is consistent with the notion that the bulk transition line ends in a critical endpoint with the continuum limit there being a $\phi^4$ theory with a diverging correlation length only in the $0^{++}$ channel.

## 1. INTRODUCTION

Studies in the early 80's of the phase diagram of fundamental–adjoint pure gauge systems,

$$S = \beta_f \sum_P [1 - \frac{1}{N} \mathrm{ReTr} U_P]$$
$$+ \beta_a \sum_P [1 - \frac{1}{N^2} \mathrm{Tr} U_P^\dagger \mathrm{Tr} U_P], \quad (1)$$

revealed a non-trivial phase structure with first order (bulk) transitions in the region of small $\beta_f$ [1,2]. In particular a line of first order transitions points towards the fundamental axis and, after terminating, extends as a roughly straight line of bulk crossovers to the fundamental axis and beyond.

In a couple of recent papers Gavai, Grady and Mathur [3] and Mathur and Gavai [4] returned to investigating the behavior of pure gauge SU(2) theory in the fundamental–adjoint plane at finite, non-zero temperature. They raised doubts about the bulk nature of the phase transition and claimed that their results were consistent with the transitions, for lattices with temporal extent $N_t = 4$, 6 and 8, to be of thermal, deconfining nature, displaced toward weak coupling with increasing $N_t$. On the transition line for a fixed $N_t$ there is then a switch from second order behavior near the fundamental axis to first order behavior at larger adjoint coupling. In a Landau Ginzburg model of the effective action in terms of Polyakov lines, Mathur [5] reported that he could reproduce the claimed behavior seen in the numerical simulations. These results, should they be confirmed, are rather unsettling, since they contradict the usual universality picture of lattice gauge theory with a second order deconfinement transition for gauge group SU(2).

Puzzled by the finding of Ref. [3], we studied the finite temperature behavior of pure gauge SU(3) theory in the fundamental–adjoint plane [6,7]. We obtained results in agreement with the usual universality picture: there is a first order bulk transition line ending at

$$(\beta_f^*, \beta_a^*) = (4.00(7), 2.06(8)). \quad (2)$$

The thermal deconfinement transition lines for fixed $N_t$ (being of first order for gauge group SU(3)) in the fundamental adjoint plane are ordered such that the thermal transition for smaller $N_t$ occurs to the left, at smaller $\beta_f$, than that for a larger $N_t$. In this order the thermal transition lines join on to the bulk transition line. The thermal transition line for $N_t = 4$ joins the bulk transition line very close to the critical endpoint. This is shown in Figure 4 of Ref. [7], reproduced here as Figure 1.

To solidify this picture, which is in agreement with the usual universality scenario, we have continued the investigation studying zero-temperature observables. We computed the string tension and the masses of some glueballs, in particular the $0^{++}$ glueball, along the thermal transition line for $N_t = 4$. For universal contin-uum behavior $\sqrt{\sigma}$ and the glueball masses should be constant along the thermal transition line for a fixed $N_t$, leading to constant ratios $T_c/\sqrt{\sigma}$ and $m_g/\sqrt{\sigma}$.

---

[*]To appear in the Proceedings of LATTICE'95, Melbourne, Australia, 11-15 July, 1995.

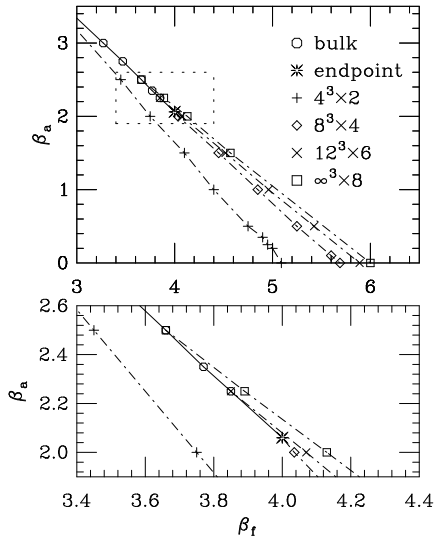

Figure 1. The phase diagram together with the thermal deconfinement transition points for $N_t = 2, 4, 6$ and $8$ from Ref. [7]. The lower plot shows an enlargement of the region around the end point of the bulk transition.

Of course, at small $N_t$, corresponding to a large lattice spacing $a$, we expect to see some deviations from this constant behavior. However, we find large deviations for $m_{0^{++}}/\sqrt{\sigma}$ as we approach the critical endpoint of the bulk transition along the $N_t = 4$ thermal transition line. The scalar glueball mass decreases dramatically; much more than what could be expected from simple scaling violations at a large lattice spacing. On the other hand, this is not really a surprise since at the critical endpoint at least one mass in the $0^{++}$ channel has to vanish. We elaborate on our findings in the next sections and then discuss the implications for the scaling behavior along the fundamental (or Wilson) axis.

## 2. OBSERVABLES AND ANALYSIS

We have made simulations of the model with action (1) along the thermal transition line for $N_t = 4$, and continued along the bulk transition line, on a $12^4$ lattice. We computed finite $T$ approximants to the potential from time-like Wilson loops constructed with 'APE'-smeared spatial links [8] to increase the overlap to the ground state potential. On and off axis spatial paths were considered with distances $R = n$, $\sqrt{2}n$, $\sqrt{3}n$ and $\sqrt{5}n$, with $n = 1, 2, \ldots$ an integer. The string tension was then extracted from the usual fit

$$V(\vec{R}) = V_0 - \frac{e}{R} + l\left(G_L(\vec{R}) - \frac{1}{R}\right) + \sigma R. \qquad (3)$$

Here $G_L(\vec{R})$ is the lattice Coulomb potential, included in the fit to take account of short distance lattice artefacts. Our fits are fully correlated $\chi^2$-fits with the correlations estimated by bootstrap, after binning to alleviate autocorrelation effects. The results of the best fits are listed in Table 1.

We also computed glueball correlation functions in the $0^{++}$, $2^{++}$ and $1^{+-}$ channel that can be built from simple plaquette operators, but built from the smeared links, already used for the computation of the potential. Not surprisingly for computations done around the critical coupling for the $N_t = 4$ thermal phase transition, we did not obtain a significant signal in the $1^{+-}$ channel and only an effective mass from time distances $t = 0/1$ in the $2^{++}$ channel – we had 500 measurements everywhere, except for $\beta_a = 2.0$, near the critical endpoint where the number was increased, as given in Table 1. In the $0^{++}$ channel we got a signal at distance $t = 1/2$ at small $\beta_a$ and out to $t = 3/4$ at $\beta_a = 2.0$. Our best results are also given in Table 1.

The quantities $\sqrt{\sigma}$ and $m_{0^{++}}$ are shown in Figure 2 plotted versus $\beta_a$. As can be seen, $\sqrt{\sigma}$ remains approximately constant along the thermal transition line for $N_t = 4$ – the errors shown in the figure are statistical only; no estimate of the error from the uncertainty in the determination of $\beta_{fc}$ has been attempted except for $\beta_a = 2.0$. There, the computation has been repeated for two nearby couplings, also listed in Table 1; the variation with $\beta_f$ becomes so rapid that the error in the determination of $\beta_{fc}$ becomes the dominating factor.

While our estimate for $m_{2^{++}}$, albeit an unreliable estimate since we had to use distances $t = 0$

| $\beta_a$ | $\beta_{fc}(N_t = 4)$ | $\beta_f$ | $L$ | $N_{meas}$ | $\sqrt{\sigma}$ | $m_{0^{++}}$ | $m_{2^{++}}(t=1)$ |
|---|---|---|---|---|---|---|---|
| 0.0 | 5.6925(2) | $5.7^a$ | | | 0.4099( 12) | 0.97( 2) | 2.39(13) |
| 0.5 | 5.25(5) | 5.25 | 12 | 500 | 0.4218( 28) | 0.93(11) | 2.29(13) |
| 1.0 | 4.85(5) | 4.85 | 12 | 500 | 0.4024( 82) | 0.78(28) | 2.45(19) |
| 1.5 | 4.45(5) | 4.45 | 12 | 500 | 0.3743( 51) | 0.56(17) | 2.13( 9) |
| 2.0 | 4.035(5) | 4.03 | 12 | 1000 | 0.555 ( 11) | 0.34( 6) | 3.11(16) |
| 2.0 | 4.035(5) | 4.035 | 12 | 2000 | 0.4725(128) | 0.20( 4) | 3.17(14) |
| 2.0 | 4.035(5) | 4.04 | 12 | 1000 | 0.3750( 24) | 0.27( 8) | 2.37( 9) |
| 2.25 | 3.8475(25) | $3.8475^b$ | 12 | 500 | 0.619 ( 18) | 0.37(10) | 3.15(37) |
| 2.25 | 3.8475(25) | $3.8475^c$ | 12 | 500 | 0.3005( 22) | 0.61( 8) | 1.59( 7) |
| 2.25 | 3.8475(25) | $3.8475^c$ | 16 | 500 | 0.2965( 19) | 0.62( 5) | 1.72( 9) |

Table 1
The results in the neighborhood of the $N_t = 4$ thermal transition line. Comments: (a) $\sqrt{\sigma}$ at $\beta_a = 0.0$ comes from [9], $m_{0^{++}}$ from [10]; we did not take their best value, but rather the effective mass from the same distance as at $\beta_a = 0.5$; $m_{2^{++}}$ is from [11]; (b) in the disordered phase and (c) in the ordered phase on the bulk transition.

and 1 to obtain enough of a signal to extract an effective mass, also remains approximately constant, $m_{0^{++}}$ shows a remarkable decrease as the critical endpoint is approached. This observed behavior suggests that the mass in the $0^{++}$ channel vanishes at the critical endpoint, thereby giving strong additional evidence for the existence of this critical endpoint, since at a critical point at least one mass, in the $0^{++}$ channel to have a rationally invariant continuum limit, must vanish.

Since no other observable seems to be dramatically affected by the critical endpoint, we conjecture that the continuum theory one would obtain there is simply the (trivial) $\phi^4$ theory. To substantiate this claim somewhat, we made a fit to the scalar mass of the form

$$m_{0^{++}} = A \left( \beta_a^* - \beta_a \right)^p \quad (4)$$

expected near a critical point. A 3-parameter fit gave $A = 0.76(11)$, $p = 0.35(20)$ and $\beta_a^* = 2.02(6)$ with a $\chi^2$ of 0.29 for 2 dof. Note that the estimate for $\beta_a^*$ is in agreement with the previous estimate (2) obtained in [7] from fits to the jump in the plaquette across the bulk transition line. Within its large error, the exponent $p$ is compatible with the mean field value 0.5 of $\phi^4$ theory, up to logarithmic corrections. Since the errors of the fit parameters are rather large we also made a fit with $\beta_a^*$ held fixed at its value 2.06 of (2). This fit gave $A = 0.71(3)$ and $p = 0.44(5)$ with a $\chi^2$ of 0.47 for 3 dof. Again, $p$ is compatible with the mean field value. Finally, a fit with $\beta_a^* = 2.06$ and $p = 0.5$ both held fixed gave $A = 0.68(1)$ with a $\chi^2$ of 1.50 for 4 dof. This last, still very acceptable fit is included in Figure 2.

## 3. IMPLICATIONS FOR SCALING

In the previous section we have corroborated the existence of a first order bulk phase transition line ending in a critical endpoint. We have provided evidence that physical observables are little affected by this phase transition line. The notable exception to this is the glueball mass in the $0^{++}$ channel which decreases as the critical endpoint is approached and vanishes there.

The influence of the critical endpoint on the $0^{++}$ glueball appears still visible in the crossover region on the fundamental (Wilson) axis. This has first been argued in Ref. [12]. The $0^{++}$ glueball is lighter in the crossover region, leading to a different scaling behavior than other observables. This can be seen in Figure 3 where we show the latest data of $T_c/\sqrt{\sigma}$ from Ref. [13] and $m_{0^{++}}/\sqrt{\sigma}$ with $\sqrt{\sigma}$ taken from Refs. [9,14] and the glueball mass from Refs. [10,15,16]. While $T_c/\sqrt{\sigma}$ stays approximately constant in the $\beta$ interval shown, $m_{0^{++}}/\sqrt{\sigma}$ decreases visibly in the crossover re-gion around $\beta = 5.7$.

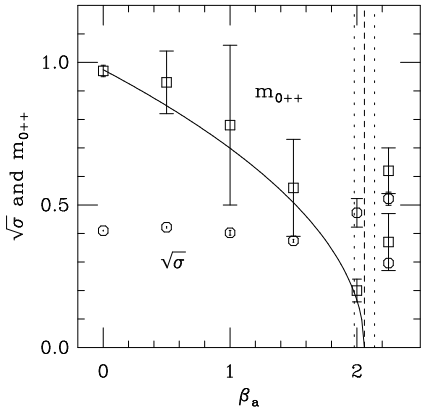

Figure 2. $\sqrt{\sigma}$ (octagons) and $m_{0^{++}}$ (squares) as a function of $\beta_a$ along the thermal transition line for $N_t = 4$. At $\beta_a = 2.25$ we show the results from both phases at the bulk transition. The dashed vertical line gives the location of the critical endpoint, (2), with the dotted lines indicating the error band. The curve is a fit to $m_{0^{++}} = A(2.06 - \beta_a)^{1/2}$.

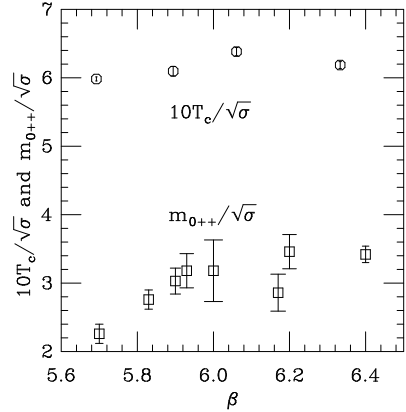

Figure 3. $10T_c/\sqrt{\sigma}$ (octagons) and $m_{0^{++}}/\sqrt{\sigma}$ (squares) as a function of $\beta$ for the fundamental (Wilson) action.

It has long been known that the scalar glueball mass scales differently in the crossover region than $T_c$ and the string tension. The scalar glueball seemed more compatible with asymptotic scaling. However, in view of our findings that the different behavior of the scalar glueball mass comes from the influence of the nearby critical endpoint this asymptotic scaling behavior appears to be accidental.

## Acknowledgements

This work was partly supported by the DOE under grants # DE-FG05-85ER250000 and # DE-FG05-92ER40742.